\documentclass[aps,prl,twocolumn,superscriptaddress]{revtex4-2}
\usepackage{graphicx}

\begin{document}

\title{Scalable generation of heralded single photons via active feed-forward switching of a fiber delay line}

\author{Xavier Barcons Planas}
\email{xavier.barconsplanas@dlr.de}
\affiliation{Institute of Space Research, German Aerospace Center, Rutherfordstr. 2, 12489 Berlin, Germany}
\affiliation{Institut für Physik, Humboldt-Universität zu Berlin, Newtonstr. 15, 12489 Berlin, Germany}

\author{Helen M. Chrzanowski}
\affiliation{Institute of Space Research, German Aerospace Center, Rutherfordstr. 2, 12489 Berlin, Germany}
\affiliation{Institut für Physik, Humboldt-Universität zu Berlin, Newtonstr. 15, 12489 Berlin, Germany}

\author{Janik Wolters}
\affiliation{Institute of Space Research, German Aerospace Center, Rutherfordstr. 2, 12489 Berlin, Germany}
\affiliation{Institut für Optik und Atomare Physik, Technische Universität Berlin, Hardenbergstr. 36, 10623 Berlin, Germany}
\affiliation{AQLS GmbH, Guerickestr. 12, 10587 Berlin, Germany}

\begin{abstract}
Quasi-deterministic single-photon generation is a key requirement for many photonic quantum technologies. Photon sources based on spontaneous parametric down-conversion (SPDC) are widely used for producing high-quality photons; however, the probabilistic nature of the process limits the generation of synchronized multi-photon states. Here, we demonstrate temporal synchronization of multiple photon-generation events using a free-space-fiber hybrid delay line with feed-forward control, enabling fast and efficient switching and scalable operation.  Narrow-band, telecom-wavelength photons compatible for fiber transmission are heralded from a monolithic cavity SPDC source and synchronized across 20 time bins. This yields a sixfold enhancement in synchronized rates and enables multi-photon synchronization, with only a marginal increase of higher-order photon-number contributions.  
\end{abstract}

\maketitle

The generation of indistinguishable single-photon Fock states at high count rates with high efficiency---synchronized to an external clock---is essential for scaling photonic quantum information applications~\cite{wang2025scalable}. Photon sources based on spontaneous parametric down-conversion (SPDC) have served as workhorses in many experiments. In this process, a high-energy photon in a nonlinear crystal spontaneously converts into a pair of lower-energy photons, historically called the signal and idler; detection of one photon heralds the presence of the other~\cite{hong1986experimental}. Nevertheless, despite several advantages---such as engineered high-purity emission~\cite{mosley2008heralded,branczyk2010engineered,kellner2026counter,barcons2026pure}, room-temperature operation, implementation simplicity, robustness, optical-circuit integrability, and replicability---the inherent probabilistic nature of the squeezing process poses a limitation on the probability of obtaining exactly one photon at the output in a given time; while increasing the mean number of photon pairs per pump pulse raises the single-photon generation probability, it simultaneously increases the likelihood of unwanted higher-order photon-number events~\cite{christ2012limits}.

To approach near-deterministic behavior with SPDC sources while maintaining a low probability of multi-photon events, one can parallelize photon creation across multiple SPDC sources through active multiplexing~\cite{meyer2020single}. Different degrees of freedom have been proposed and demonstrated for multiplexing: time~\cite{pittman2002single}, space~\cite{migdall2002tailoring,ma2011experimental}, frequency~\cite{grimau2017heralded,joshi2018frequency,hiemstra2020pure}, and hybrid approaches~\cite{mendoza2016active}. Temporal multiplexing is particularly attractive because it scales favorably regarding physical resources and alignment effort. Switchable delay lines have been studied for temporal multiplexing, in which the presence of a photon is heralded, stored in an optical cavity via active switching, and released into a predefined time bin, enabling discrete temporal delays~\cite{pittman2002single} and quasi-deterministic single-photon generation with high efficiency~\cite{kaneda2019high}. Variations of this architecture have been implemented for quantum storage~\cite{pittman2002cyclical,kaneda2017quantum,evans2023experimental}, generation of Bell states~\cite{pegoraro2024demonstration} and GHZ states~\cite{meyer2022scalable}, quantum simulation via quantum walks~\cite{schreiber20122d}, and generation of three-dimensional cluster states to demonstrate quantum advantage~\cite{madsen2022quantum}, among others. Alternatively, one can implement time multiplexing in a network of several delay lines~\cite{mower2011efficient}, thereby reducing the number of required delay cycles and associated losses. This has been implemented in full fiber-based setups~\cite{xiong2016active,lee2019relative}, though fiber-optic switches add considerable loss to each multiplexed mode, and their low switching speed limits the clock rates. Consequently, a free-space-fiber hybrid delay line would profit from Pockels cells---benefiting from their fast electro-optic switching and close-to-unity transmission---as well as from the flexibility and scalability of fiber optics. This motivates the design of an actively switched low-loss fiber delay line for synchronizing SPDC photons.

\begin{figure*}[t]
\centering
\includegraphics[width=\textwidth]{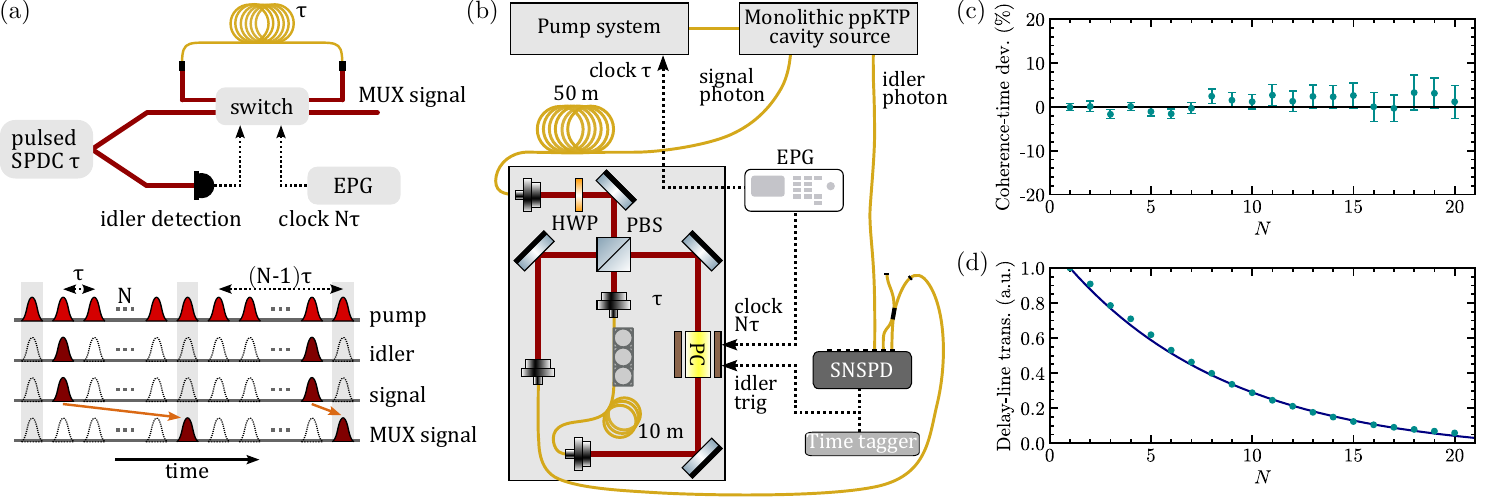}
\caption{(a) Concept of photon synchronization via multiplexing (MUX) using a free-space-fiber hybrid switchable delay line. Telecom and narrow-band photons heralded from a photon-pair source are stored in the system using a free-space electro-optic switch and released at clock-synchronized time bins. (b) Experimental setup based on the concept presented in (a), in which the optical switch is implemented by combining a polarizing beamsplitter with a Pockels cell. Narrow-band telecom photons are generated via SPDC in a monolithic cavity~\cite{barcons2026pure} and synchronized with the fiber delay line. HWP: half-wave plate; PBS: polarizing beamsplitter; PC: Pockels cell; EPG: electrical pulse generator; SNSPD: superconducting nanowire single-photon detector. (c) Relative deviation of the measured multiplexed-photon coherence time from its weighted mean versus the number of delay cycles, with each cycle corresponding to 50~ns of fiber propagation delay. The error bars represent one standard deviation. (d) Measured delay-line transmission as a function of the number of delay cycles, together with the corresponding exponential fit. The inferred cycle efficiency is $\eta_c$=0.89.}
\label{fig:muxconcept}
\end{figure*}

In this letter, we demonstrate active synchronization of heralded photons by a delay line with an optical fiber and free-space switching for scalable temporal multiplexing. We employ a monolithic-cavity heralded SPDC single-photon source~\cite{barcons2026pure}, which produces narrow-band, telecom C-band photons that are highly suitable for fiber-optic transmission and largely insensitive to dispersion effects. The delay line exhibits a high cycle transmission exceeding 89\%, including fiber coupling and switching losses. Via synchronization across up to 20 time bins, we achieve 6-, 29-, and 180-fold improvements in the detected rates for 1, 2, and 3 consecutively heralded photons, respectively, relative to the non-multiplexed case. The multi-photon contamination of the state purity degrades by a factor of 1.5 through all settings.

The conceptual idea of our approach is depicted in Fig.~\ref{fig:muxconcept}(a). It adapts the original free-space storage-loop multiplexing scheme demonstrated by T. B. Pittman, B. C. Jacobs, and J. D. Franson~\cite{pittman2002single} to synchronize the emission of photons to a fixed external clock, with additional flexibility to different source repetition rates and the potential for scalability in fiber setups. A telecom narrow-band photon-pair source---e.g., based on SPDC in a monolithic cavity~\cite{barcons2026pure}---is pumped with period $\tau$. A successful heralding event triggers a free-space low-loss switch through feed-forward operation, which stores the corresponding signal photon of the pair in a fiber delay line with temporal delay $\tau$. Here it is ideally stored until an external periodic clock signal with a period $N\tau$ releases it from the delay line. This setup emits photons temporally synchronized with the external clock and allows $N$ attempts of photon generation per clock cycle, which enhances the clock-synchronized photon generation efficiency. A critical requirement for this fiber setup is the use of photons that have low fiber attenuation and chromatic dispersion, which implies the need for photons at telecom wavelengths and narrow bandwidths. Dispersion effects are practically negligible for narrow-band sources producing $\sim$1~ns photons in 1~km of single-mode fiber ($\sim$17~ps/(nm~km) at 1550~nm); in contrast, for standard sources generating, for instance, $\sim$3~ps photons, dispersion becomes significant and broadens the pulse duration to $\sim$20~ps, requiring additional dispersion-compensation fibers, which add losses and photon distinguishability~\cite{lee2019relative}.

The enhancement of the clock-synchronized rate by multiplexing can be mathematically described as follows. The probability of generating $k$ photon pairs in an SPDC source follows the Bose-Einstein distribution $P_k = (\eta\mu)^k/(\eta\mu+1)^{k+1}$, where $\mu$ is the mean number of photon pairs per pulse and $\eta$ the system transmission and detection efficiency for $2k$ photons. The probability of generating a single photon pair $P_1$ is limited due to the non-negligible probability of producing either zero or multiple pairs, with an upper bound of $P_1=25\%$ for $\eta=1$ and ideal photon-number-resolved detection~\cite{christ2012limits}. With multiplexing over $N$ time bins, $P_1$ is no longer limited and can reach up to $P_1^{\text{MUX}}=1-(1-P_1)^N$ in the lossless case. Therefore, for a fiber delay line with cycle efficiency $\eta_c$ and a clock rate $R_{\text{sync}}$, the multiplexed clock-synchronized $n$-photon generation rate is 
\begin{equation}
    R_n(N) = R_{\text{sync}}\left[1-\prod_{r=1}^N{(1-P_1\eta_c^r)}\right]^n,
\label{eq:syncrate}
\end{equation}
which corresponds to the coincidence rate of $n$ consecutively heralded photons accounting for optical losses ($2n$ photons including the heralding idler photons). The $n$-photon on-demand generation efficiency is thus defined by $\varepsilon_n(N) = R_n(N)/R_{\text{sync}}$, and the $n$-photon rate enhancement by synchronization is $E_n(N)=R_n(N)/R_n(1)$. The non-multiplexed rate is given by $R_n(1)$ and corresponds to the setting with an inactive switch, in which, in the setup presented here, the photons still require one cycle through the delay line. These expressions hold only in the low gain regime, where multi-pair contributions are minimized.

The experimental setup is illustrated in Fig.~\ref{fig:muxconcept}(b). The photon-pair source utilizes pulsed SPDC within a monolithic ppKTP cavity~\cite{barcons2026pure}. It can herald photons into a well-defined cavity spectral mode of 168~MHz bandwidth (1.3~ns coherence length) at a 1540~nm wavelength with measured 96(3)\% spectral purity. Consequently, these photons are highly suitable for fiber transmission, offering both low loss and negligible chromatic dispersion. Figure~\ref{fig:muxconcept}(c) shows the measured coherence time of the multiplexed photon (represented as a relative deviation from the weighted mean) as a function of the number of fiber delays, indicating the insignificant broadening through 200~m of fiber. Additionally, the source generates photons into a well-defined spatial mode that can be efficiently coupled into single-mode fibers with over 90\% efficiency. The source is pumped at a repetition rate of $\mathord{1/\tau\!\sim\!19.42}$~MHz, chosen to match the temporal delay $\tau$ of the delay line. By detection of the idler photon with a superconducting nanowire single-photon detector (SNSPD), the presence of the signal photon is heralded. The signal photon propagates through 50~m of fiber, allowing for feedforward of the electronic signal to the switch before the photon enters the delay line.

The switchable delay line consists of a 10~m fiber combined with a free-space section, setting the total delay cycle to $\tau=51.5$~ns. The switching mechanism is implemented in free space with a high-speed Pockels cell (PC) and a polarizing beamsplitter to store the photons in the system or release them on demand. The setup comprises an anti-reflection-coated single-mode fiber with both ends mounted on a three-axis flexure stage, with 11~mm focal length plano-convex aspheric lenses ensuring controlled mode-matching from fiber to fiber. The fiber is guided through a manual fiber polarization controller to compensate for polarization changes caused by birefringence. The system was carefully optimized, maximizing the transmission through all the components and fiber couplings, achieving an in-line fiber coupling efficiency as high as 94\%. The transmission of the delay line as a function of the number of cycles is presented in Fig.~\ref{fig:muxconcept}(d). A cycle  transmission of $\eta_c=89\%$ was achieved, accounting for fiber-coupling losses and imperfect extinction of the polarization optics, corresponding to a $1/e$ lifetime of 430~ns, namely, approximately 8 time bins. The photons are coupled into and out of the delay line via optical fibers with efficiencies over 89\% and 80\%, respectively. Additionally, about 2\% of the photons fail to enter the delay line due to imperfect linear polarization.

Storage and release operate as follows. When a photon pair is generated, detection of the idler triggers the PC via feed-forward control, which rotates the linear polarization of the signal photon by $90^{\circ}$ to store it in the delay line. Specifically, the SNSPD voltage pulse produced after the idler detection is recorded with the time-tagger for subsequent data analysis and delayed to trigger the PC to store the photon. This signal is combined with a second electrical trigger from an electrical pulse generator, phase-locked to the pump system, to periodically release the stored photon. The two signals are then amplified and jointly used to drive the PC. Photon release is fixed by the electrical pulse generator to a clock rate of $R_{\text{sync}}=1/(20\tau)\!\sim\!0.97$~MHz, permitting up to 20 multiplexed time bins per synchronization cycle. Subsequently, the released signal photons are directed into a Hanbury Brown and Twiss interferometer, detected with SNSPDs (with manufacturer-reported detector efficiencies of 87--93\%) and recorded with a time tagger.

\begin{figure}[t]
\centering
\includegraphics[width=\linewidth]{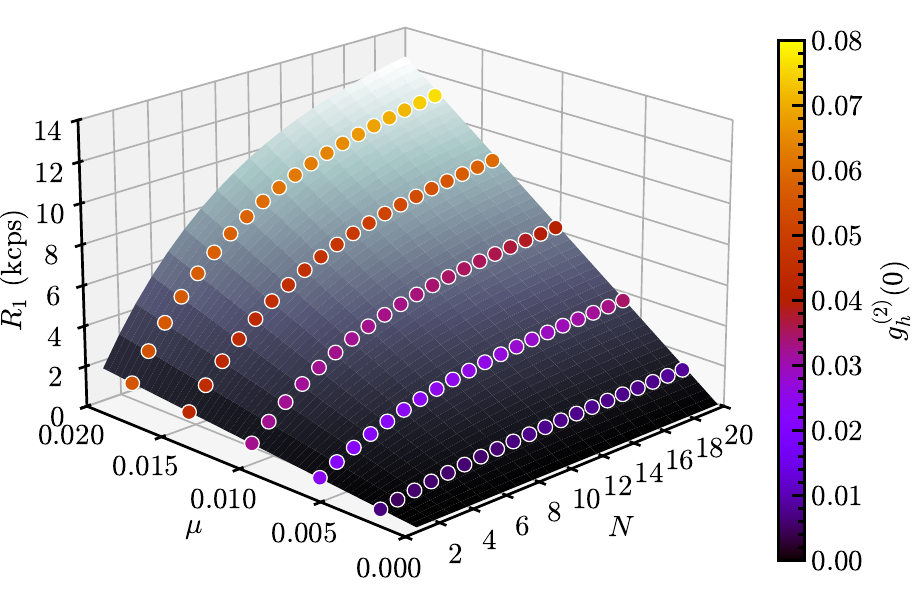}
\caption{Measured 1-photon detection rate $R_1$ as a function of the mean photon number per pulse $\mu$ and the number of multiplexed time bins $N$. The marker color encodes the measured higher-order photon-number contributions. Both quantities are defined as cumulative over $N$ for a fixed $\mu$ ($g_h^{(2)}(0)$ as a weighted average), indicating the marginal increase of $g_h^{(2)}(0)$ with multiplexing, whereas the multiplexed rate is enhanced by approximately one order of magnitude. The surface points are described by the theory (Eq.~\ref{eq:syncrate} for $n=1$).}
\label{fig:rates_vs_numroundtrips}
\end{figure}

The arrival times of the signal and idler photons were registered for both the multiplexed and non-multiplexed cases---namely, with the Pockels cell off---for various average pump powers $\mathcal{P}$, and thus for different mean photon numbers $\mu$, specifically $(\mathcal{P} \text{(mW)},\mu,\eta\mu\eta_c)=\{$$(2.3,0.002,0.0002)$, $(6.0,0.006,0.0007)$, $(10.3,0.010,0.0011)$, $(14.2,0.014,0.0015)$, $(18.2,0.018,0.0019)\}$. Here, $\eta\mu\eta_c$ denotes the effective transmitted and detected $\mu$ for $N=1$. A time-tag analysis of the multiplexed 1-photon rate $R_1$ and heralded second-order autocorrelation $g_h^{(2)}(0)$ was performed based on the mean photon number per pulse and the number of multiplexed time bins. The results are shown in Fig.~\ref{fig:rates_vs_numroundtrips}. $R_1$ increases with the number of multiplexed time bins and saturates around $N=20$ due to the exponentially decaying delay-line transmission, reaching about an order of magnitude increase, while $g_h^{(2)}(0)$ shows only a moderate rise.

\begin{figure}[t]
\centering
\includegraphics[width=\linewidth]{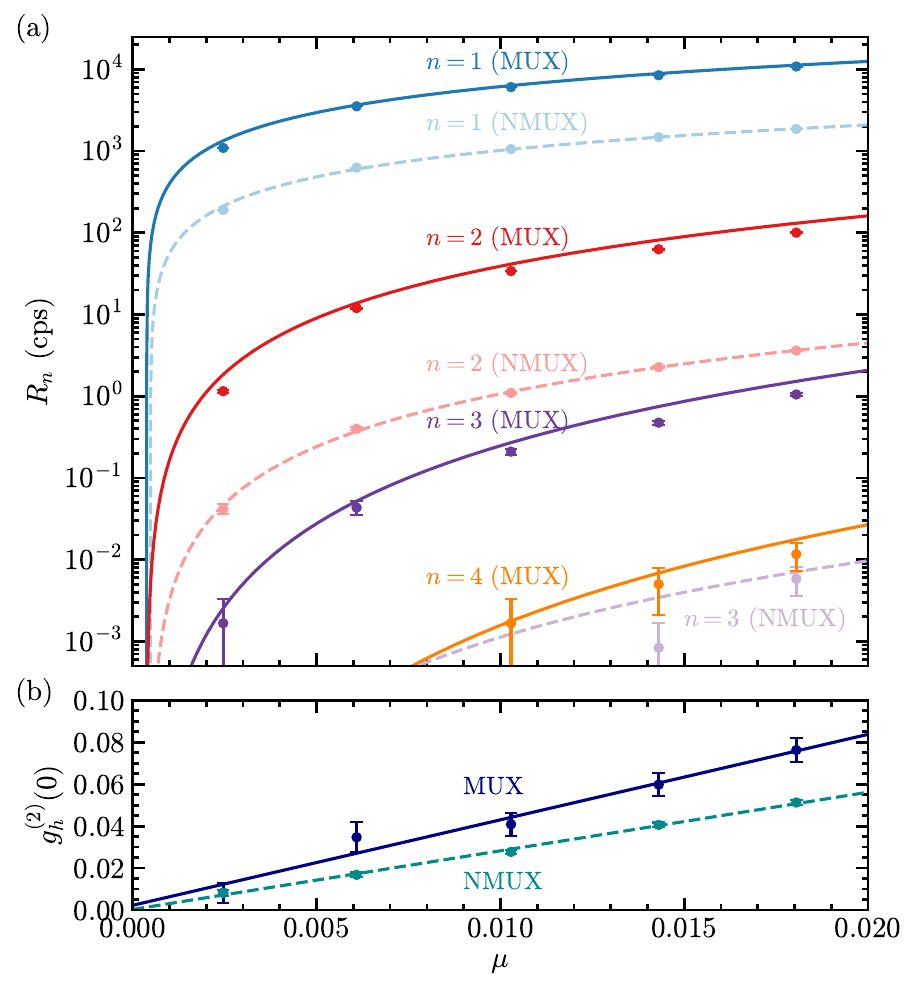}
\caption{(a) Measured $n$-photon detection rates $R_n$, and (b) measured multi-photon contamination $g_h^{(2)}(0)$ as a function of the mean photon number per pulse $\mu$ for multiplexed (MUX, solid dark lines) and non-multiplexed (NMUX, dashed light lines) configurations. The theoretically predicted $n$-photon rates (Eq.~\ref{eq:syncrate}) are overlaid in (a), and a linear regression is applied to the data points in (b). The integration time for each data point is 10~(20)~min for MUX (NMUX) configurations. The error bars are derived from Poissonian photon counting statistics.}
\label{fig:rates_vs_power}
\end{figure}

A subsequent time-tag analysis was carried out to determine the clock-synchronized rates $R_n$ of $n=1$, 2, 3, and 4 consecutively heralded photons for different pump powers, together with their associated $g_h^{(2)}(0)$, which is calculated as a weighted average over the 20 time bins. The results are presented in Fig.~\ref{fig:rates_vs_power}, along with the predicted theoretical curves from Eq.~\ref{eq:syncrate}. As expected from the theory, $R_n$ scales exponentially with $n$. Note that each $n$-photon rate $R_n$ is actually comprised of the detection of $2n$ photons, since each signal photon is heralded to trigger the electronics logic. A modest 1-photon multiplexed rate $R_1$ on the order of $10^3~\mathrm{s^{-1}\,mW^{-1}}$ was achieved, representing roughly a sixfold improvement over the non-multiplexed case, though the performance is largely constrained by the limited $P_1$ of the cavity source and the low clock rate forced by the triggering requirements of the Pockels cell. For $n>1$ the impact of multiplexing is highly pronounced. For instance, a 3-photon multiplexed rate $R_3$ on the order of $1~\mathrm{s^{-1}}$ was observed at a pump power of 18.2~mW ($\mu=0.018)$ with $g_h^{(2)}(0)=0.076(6)$, corresponding to an increase by two orders of magnitude over the non-multiplexed setting. Additionally, the 4-photon multiplexed rate $R_4$ exceeded the 3-photon non-multiplexed rate. Overall, multiplexing yields measured rate enhancements of $E_1\!\sim\!6$, $E_2\!\sim\!29$, and $E_3\!\sim\!180$ relative to the non-multiplexed case. A 4-photon rate enhancement of $E_4\!\sim\!1080$ is expected, which could not be experimentally verified as no non-multiplexed 4-photon events were measured in the integration time of 20~min. While the rates are substantially enhanced, the $g_h^{(2)}(0)$ degrades by a factor of 1.5 in all cases. The higher-order photon-number contributions increase and are more prominent in the multiplexed case, as higher-order photon numbers have a higher survival bias; that is, they are more likely to pass through a system with losses and still contribute to detection events.

In conclusion, we have demonstrated a scalable approach to synchronize multiple heralded photon-generation events. By employing a narrow-band photon-pair source, we enable free-space-fiber hybrid multiplexing with flexibility and potential for scalability. Efficient linear optical quantum computation requires the product of detector and source efficiencies to exceed $2/3$~\cite{varnava2008good}, and state-of-the-art photon sources have demonstrated increasingly high efficiencies towards it: F. Kaneda et al.~\cite{kaneda2019high} realized a free-space-multiplexed SPDC source with $\sim$67\% single-photon collection efficiency with $g_h^{(2)}(0)$ of $\sim$0.27, while deterministic emitters achieved $\sim$43\% end-to-end efficiency in cavity-coupled single atoms~\cite{thomas2022efficient} and $\sim$71\% collection efficiency in a quantum dot deterministically coupled to a tunable open cavity with $g_h^{(2)}(0)$ of $\sim$0.02~\cite{ding2025high}, exceeding the scalability threshold; however, scaling to multiple identical quantum dots remains a challenge. Despite the fundamental probabilistic nature of SPDC, the synchronized photon rate in our approach is primarily limited by engineering constraints: the usable photon rate before multiplexing is reduced by the intrinsic low spectral bandwidth of cavity-enhanced type-II SPDC sources  and imperfect heralding efficiencies; moreover, cumulative optical losses in the delay line from fiber couplings and imperfect polarization rotation by the Pockels cell, as well as timing electronics, add additional technical limits. We therefore expect that optimization of the delay-line transmission---for example, adding more degrees of freedom for fiber coupling and polarization control---or utilizing a near-optimal SPDC source will boost efficiency and allow to achieve higher rates. For instance, a similar source with $\sim$90\% heralding efficiencies, and a delay line with $\sim$95\% cycle transmission, could achieve $\sim$67\% generation efficiency for $\mu\!\sim$0.2, reaching therefore the threshold for efficient quantum computing. Furthermore, highly efficient photon-number-resolving detectors could suppress the multi-photon contamination of the state.

Crucially, this approach scales favorably: synchronized rates improve several orders of magnitude with only a marginal increase in $g_h^{(2)}(0)$, enabling either higher photon probabilities or equivalent rates at lower pump power and thus a lower $g_h^{(2)}(0)$. More generally, such feed-forward operations are key components for photonic quantum technologies. In future experiments, the $n$ synchronized photons within a single fiber could be routed into $n$ separate fibers by employing a temporal-to-spatial demultiplexer~\cite{anton2019interfacing} and serve as the resource for many interferometric applications, e.g., for heralded generation of GHZ states~\cite{wiesner2024influence} or quantum computing in a photonic quantum processor~\cite{maring2024versatile}. These improvements are essential for scaling multi-photon quantum optics experiments.

\textbf{Funding.} Bundesministerium für Forschung, Technologie und Raumfahrt (13N15870).

\textbf{Acknowledgment.} The authors thank Tobias Martens for assistance with the initial experimental setup.

\textbf{Disclosures.} The authors declare no conflicts of interest.

\textbf{Data Availability Statement.} Data underlying the results presented in this paper are not publicly available at this time but may be obtained from the authors upon reasonable request.

\bibliographystyle{apsrev4-2} 
\bibliography{references}

\end{document}